\documentclass[sn-mathphys,Numbered]{sn-jnl}

\usepackage{graphicx}%
\usepackage{multirow}%
\usepackage{amsmath,amssymb,amsfonts}%
\usepackage{amsthm}%
\usepackage{mathrsfs}%
\usepackage[title]{appendix}%
\usepackage{xcolor}%
\usepackage{textcomp}%
\usepackage{manyfoot}%
\usepackage{booktabs}%
\usepackage{algorithm}%
\usepackage{algorithmicx}%
\usepackage{algpseudocode}%
\usepackage{listings}%

\begin{document}

\title[LV]{On the combinatorics of Lotka-Volterra equations}


\author*[1]{\fnm{Francesco} \sur{Caravelli}}\email{caravelli@lanl.gov}

\author[2]{\fnm{Yen Ting} \sur{Lin}}\email{yentinglin@lanl.gov}
\equalcont{These authors contributed equally to this work.}

\affil*[1]{\orgdiv{T-4}, \orgname{T-Division, Los Alamos National Laboratory},\\ \orgaddress{\street{Bikini Atoll rd}, \city{Los Alamos}, \postcode{87544}, \state{NM}, \country{USA}}}

\affil[2]{\orgdiv{CCS-3}, \orgname{CCS-Division, Los Alamos National Laboratory},\\ \orgaddress{\street{Bikini Atoll rd}, \city{Los Alamos}, \postcode{87544}, \state{NM}, \country{USA}}}


\abstract{
We study an approach to obtaining the exact formal solution of the 2-species Lotka -
Volterra equation based on combinatorics and generating functions. By employing a combination
of Carleman linearization and Mori-Zwanzig reduction techniques, we transform the nonlinear equations into
a linear system, allowing for the derivation of a formal solution.
The Mori-Zwanzig reduction
reduces to an expansion which we show can be interpreted as a directed and weighted lattice path
walk, which we use to obtain a representation of the system dynamics as walks of fixed length. The exact solution is then shown to be dependent on the generator of weighted walks. We show that the generator can be obtained by the solution of PDE which in turn is equivalent to a particular Koopman evolution of nonlinear observables.}




\maketitle
\clearpage
\tableofcontents
\section{Introduction}

 The two species Lotka-Volterra equations are the archetypal nonlinear equations, exhibiting both laminar and oscillatory behavior. These are also known as predator-prey equations and are a pair of coupled differential equations used to model the interactions between two species in an ecosystem. Developed independently by Alfred J. Lotka and Vito Volterra in the early 20th century  \cite{lotka1925elements,lotka1926fluctuations,lotka1926contribution,volterra1926variations}, these equations provide a mathematical framework for understanding the dynamics of predator-prey relationships and to model the nonlinearity of co-dependent species. This model has experienced some renewed interest in later years because random multi-species models exhibit marginally stable equilibria \cite{bunin}. The techniques used to study these problems come from disordered systems, which have shed some light on a century-old problem.  While the LV equations have been extensively studied, and various analytical and numerical methods have been employed to find solutions and analyze their behavior, we believe that the results and combinatorial structure presented in this paper have never been explored in the past \cite{bookpopdyn,bookpopdyn2}.

The Lotka-Volterra equations assume two main populations: the predator and the prey. They describe how the populations of these species change over time based on their interactions and the availability of resources. In the most general form, the equations are as follows:
\begin{eqnarray}
\frac{dx}{dt} &=& \alpha x + \beta xy  \\
\frac{dy}{dt} &=& \gamma y + \delta xy
\end{eqnarray}
where $x$ and $y$ represent the real-valued population densities of the prey (e.g. rabbits) and predator (e.g. wolves) species, respectively.
The quantities $dx/dt$ and $dy/dt$ represent the rates of change of the population densities over time. The parameters $\alpha>0$ and $\gamma<0$ represent the per-capita growth rate of the prey and predator population in the absence of predation, respectively (e.g. rabbits reproduce indefinitely in the absence of wolves, while wolves die off in the absence of rabbits). The nonlinear terms are controlled by the parameters
$\beta$
and $\delta$: $\beta<0$ and $\delta>0$ represent the rate of change of prey and predator population (per unit of prey population, per unit predator population) due to predation.  Above, one often assumes the ecologically meaningful parameter regime: $\alpha>0$, $\beta<0$, $\gamma<0$, and $\delta>0$, where oscillatory dynamics can occur as a result of feedback, instead of unstable growth or complete decay.  This paper is predominantly focused on the two species case, but we will later discuss some features of higher dimensional LVs as well in view of the combinatorial analysis of this paper.

The LV equations illustrate the complex interplay between predators and prey in an ecosystem. They demonstrate how fluctuations in one population can influence the dynamics of the other population. These equations have been used to study various ecological systems, such as the interaction between hares and lynx \cite{haresLynx} or bacteria and Bacteriophages \cite{maslovPopulationCyclesSpecies2017}, and provide insights into population dynamics, cycles, and stability \cite{strogatz2000nonlinear}. While the LV equations offer a simplified representation of predator-prey dynamics, they provide a valuable tool for understanding the fundamental principles of ecological interactions and have contributed to developing more sophisticated ecological models (e.g., \cite{Martcheva2015,RosenzweigMacArthur}). This is also why understanding the structure underlying the two-species model can provide insights into more complex ecological networks \cite{grilli}.


 The present manuscript is the result of an attempt to highlight certain features of coupled and nonlinear ODE that have been neglected. Using 
 the Lotka-Volterra (LV) equations as our model, we study the relationship between quantities that are combinatorial in nature, such as generating functions of weighted walks, the Koopman operator, and how these two are related formal exact solution of the two species LV.

The relationship between Koopman evolution \cite{Koopman255,Koopman315,kamb20time}, Mori-Zwanzig methods \cite{mori1965transport,zwanzig1973nonlinear,zwanzig2001nonequilibrium}, and ordinary differential equations (ODEs) lies at the intersection of dynamical systems theory and statistical mechanics \cite{evans2008StatisticalMechanicsNonequilibrium}. 
The Koopman operator is a mathematical tool used in the study of dynamical systems. It describes the evolution of observables (functions) over time without explicitly solving the underlying ODEs. The Koopman operator provides a linear representation of the dynamics, enabling the analysis of complex systems often through spectral methods and linear algebra techniques. It is particularly useful for systems with high-dimensional state spaces. On the other hand, Mori-Zwanzig methods are a set of mathematical techniques used in statistical mechanics to derive effective equations of motion for a system with many degrees of freedom \cite{espanol,darveComputingGeneralizedLangevin2009,Lin2021,li2015incorporation,wang2019implicit,parish2017non,falkena2019DerivationDelayEquation}. The main idea is to systematically eliminate the ``fast" degrees of freedom (unresolved variables) to obtain a reduced model that only considers the ``slow" (resolved) or relevant variables. This reduced model is often in the form of ODEs and captures the long-term behavior of the system while discarding fast oscillations or fluctuations.
The Koopman operator and Mori-Zwanzig methods are related in the context of understanding the dynamics of complex systems \cite{linDatadrivenModelReduction2021,lin2021DataDrivenLearningMori}. The Koopman operator provides an abstract and linear perspective on the system's evolution \cite{mauroy2016GlobalStabilityAnalysisa,wanner2020robust}, while Mori-Zwanzig methods focus on obtaining reduced models for high-dimensional systems in terms of ODEs.  The Koopman operator can be used to study the dynamics of observables, and by exploiting its spectral properties, one can identify relevant slow modes that capture the system's long-term behavior. The Koopman operator has been used extensively in fluid dynamics \cite{Arbabi17ergodic}, where it has been linked to the Dynamic Mode Decomposition \cite{schmid2010DMD,schmid2011applications}. By constructing a reduced model using Mori-Zwanzig methods based on these slow modes, one can effectively obtain a system of ODEs that approximates the original dynamics. 

As we show in this paper for two sets of differential equations, the additional insight that we gain in this manuscript is that the Koopman evolution is indeed also the same operator that generates certain weighted and directed lattice walks, which directly enter into the formal solution of these equations. These lattice walks can be thought to be associated with monomials in the initial conditions.

The paper is organized as follows. We first introduce in section 2 the two key techniques that we use to formally solve the equations, given by a Carleman linearization \cite{carleman1932,carleman} first in Sec. \ref{sec:21}, followed by a Mori-Zwanzig reduction in Sec. \ref{sec:22}. We then first solve the well-known equation $\dot x=x^2$ in Sec \ref{sec:31}-\ref{sec:32}-\ref{sec:33}, showing that the solution has an interpretation in terms of a one-dimensional lattice walk, and show that the solution depends on the generator of lattice walks. We derive a closed-form PDE for these generators in Sec. \ref{sec:34}, with an analysis of the utility of this representation in Sec. \ref{sec:35}. We then focus on the Lotka-Volterra equations in Sec. \ref{sec:4}. We introduce the variables used for the Carleman linearization in Sec. \ref{sec:41}, and perform the Mori-Zwanzig reduction in \ref{sec:42}, introducing the expression of the formal solution of the LV. In Sec. \ref{sec:43} we provide the interpretation of the formal solution in terms of weighted lattice walks, we introduce the lattice coefficients, showing that we can in principle use Monte Carlo techniques to obtain the solution in Sec. \ref{sec:44}. In Sec. \ref{sec:45} we show that the solution can be analyzed in terms of the generating function of lattice walks, and provide a closed-form quasi-linear PDE for the solution of the memory kernel of the solution, which is the generating function lattice walks, and analyzing the formal solution of the PDE in terms of the Lagrange-Charpit method.  In Sec. \ref{sec:46} we provide a brief analysis of the N-species Lotka-Volterra equations. Conclusions follow.

\section{Formal solution method} \label{sec:2}

We begin by providing the main ingredients which allow to formally write the solution of the LV equations. We first discuss Carleman linearization, and then discuss the Mori-Zwanzig formalism.

\subsection{Linearization} \label{sec:21}
Carleman linearization \cite{carleman1932} is a powerful technique used to transform nonlinear differential equations into linear ones \cite{carleman}, facilitating their analysis and solution. This method involves introducing additional variables and expressing the original nonlinear equations as a linear system with respect to these new variables. By doing so, Carleman linearization enables the use of well-established linear techniques to investigate the behavior and properties of the system. In this paper, we utilize Carleman linearization to obtain lattice path expansions, which serve as the foundation for deriving the exact formal solution of the 2-species Lotka-Volterra equation.
Consider a system of differential equations of the form
\begin{equation}
\frac{d\vec x}{dt}=\vec f(\vec x).\label{eq:systemof}
\end{equation}
We can linearize this system by considering a set of observables given by the terms of observables of the form $r_{\vec a}=\prod_i x_i^{a_i}$ with $a_i\in \mathbb N$. Then, from eqn. (\ref{eq:systemof}), expanding $f_i(\vec x)$ (assuming they are analytical), one also writes the infinite chain of equations in terms of $r_{i}$. If $f$ is time-invariant, then the result equation $\dot \vec r=A \vec r$ is also time-invariant, and formally the solution for this system is known. However, the chain of equations is infinite-dimensional. We use instead a variant of this technique, which is the Mori-Zwanzig formalism. However, the key property of quadratic equations such as Lotka-Volterra, as we show below, is that they are amenable to a combinatorial treatment. 

\subsection{The Mori-Zwanzig formalism for linear system} \label{sec:22}
Let us consider the simplest example of Mori-Zwanzig coarse-graining \cite{espanol}. We have a linear ODE system of the form
\begin{eqnarray}
    \frac{d}{dt} \vec x=A \vec x
\end{eqnarray}
with $\vec x(0)=\vec x_0$. We wish to express the system's dynamics in a generalized Langevin formalism. We then assume that our observables are a subset of the vector $\vec x$, e.g.
\begin{eqnarray}
    \vec x(t)=\begin{pmatrix}
        \vec y(t) \\
        \vec z(t)
    \end{pmatrix}
\end{eqnarray}
with $\vec x\in \mathcal R^N$, while $\vec y\in \mathcal R^{N-m}$ and $\vec z\in \mathcal R^{m}$. We partition the matrix $A$ in blocks, so that
\begin{eqnarray}
    A=\begin{pmatrix} A_{rr} & A_{ru} \\
    A_{ur} & A_{uu}
    \end{pmatrix}
\end{eqnarray}
where $A_{rr}$ is $(N-m) \times (N-m)$, $A_{ru}$ is $N\times (m \times m)$, $A_{ur}$ is $m\times (N-m)$ and $A_{uu}$ is of size $(m\times m)$. Then, we can write
\begin{eqnarray}
\frac{dy}{dt}&=& A_{rr} \vec y+  A_{ru} \vec z\\
\frac{dz}{dt}&=& A_{ur} \vec y+A_{uu}\vec z.
\end{eqnarray}
In the blocks above, $r$ stands for resolved variables, while $u$ stands for unresolved variables.
Our observables, thus the resolved variables, are the $y$ components. We thus write the formal expression for $z$, which we assume to be unresolved, e.g.
\begin{equation}
z(t)=e^{A_{uu} t} \vec z_0+\int_0^t ds\ e^{A_{uu} (t-s)} A_{ur} \vec y(s).
\end{equation}
We now plug this expression into the first set of equations, obtaining
\begin{eqnarray}
    \frac{dy}{dt}&=& A_{rr} \vec y+\int_0^t ds\ A_{ru} e^{A_{uu} (t-s)} A_{ur} \vec y(s)+A_{ru} e^{A_{uu} t} \vec z_0 \label{eq:morizwanziglin}
\end{eqnarray}
We then identify the noise term $F(t)=A_{ru} e^{A_{uu} t} \vec z_0$ and the kernel operator, given by
\begin{eqnarray}
    K(t-s)=A_{ru} e^{A_{uu} (t-s)} A_{ur}.
\end{eqnarray}
from which then we obtain
\begin{eqnarray}
    \frac{d\vec y}{dt}&=& A_{rr} \vec y+\int_0^t ds\ K(t-s) \vec y(s)+\vec F(t),
\end{eqnarray}
which is in the generalized Langevin equation. Thus, our problem reduces to a simpler problem to analyze than the generalized Langevin equation, and the memory kernel can be written down explicitly.

\section{Simple example: $\dot x=x^2$} \label{sec:3}
The techniques introduced in the previous sections can be showcased on a particular ODE, whose solution has many of the key characteristics of the LV equations.
In particular, after having applied the Carleman linearization and the Mori-Zwanzig formalism, we will see that the solution of this equation obtained via power series resummation can be shown to be connected to the generator of (weighted) lattice walks in one dimension.
We consider the nonlinear ODE
\begin{eqnarray}
    \frac{dx}{dt}=x^2,\ x(0)=x_0.\label{eq:n1d}
\end{eqnarray}
The solution is given by $x(t)=x_0(1-x_0 t)^{-1}$. To see this,  we can write
\begin{equation}
\frac{1}{x^2} \frac{d}{dt}  x = \frac{d}{dt} (-\frac{1}{x})=1    
\end{equation}
and thus
\begin{eqnarray}
    \frac{1}{x_0}-\frac{1}{x(t)}=t
\end{eqnarray}
which, inverting, gives
\begin{eqnarray}
    x(t)=\frac{x_0}{1-x_0 t}.
\end{eqnarray}

\subsection{Carleman linearization and Mori-Zwanzig reduction} \label{sec:31}
We now use the Carleman linearization, we have that the observables are the Taylor powers, given by
\begin{eqnarray}
    r_k=x^k.
\end{eqnarray}
It is not hard to see that the equation above can be written in the form
\begin{eqnarray}
    \frac{d}{dt} \vec r=A \vec r
\end{eqnarray}
with $A_{ij}=  i\delta_{i+1,j}$. We will treat $r_1$ as the resolved observable, and the rest $r_k$, $k\ge 2$ as the unresolved observables. 

We now use eqn. (\ref{eq:morizwanziglin}).
First, note that $A_{rr}=A_{ur}=0$. Then, our equation for the observable $r_1$ reads
\begin{eqnarray}
    \frac{dr_1}{dt}=\frac{dx}{dt}=A_{ru} e^{A_{uu} t} \vec z_0.
\end{eqnarray}
and then
\begin{eqnarray}
    x(t)=x_0+A_{ru}\int^t ds\ e^{A_{uu} s} \vec z_0.
\end{eqnarray}
In the equation above, $\vec z_0=(x_0^2,x_0^3,\cdots)$. The vector $A_{ru}$ is infinite and looks like $A_{ru}=(1, 0,\cdots, 0)$ while $A_{uu}$ is a square super-diagonal matrix, where the super diagonal is infinite, and given by $(2,3,4, \cdots).$
We must then calculate exp$(A_{uu} t)$. 
\subsection{Series summation} \label{sec:32}
We use the expression
\begin{eqnarray}
    e^{A_{uu} t}=\sum_{k=0}^\infty \frac{t^k}{k!} A_{uu}^k,
\end{eqnarray}
for which we need to calculate an expression for the first row of the matrix, as 
\begin{eqnarray}
    A_{ru} \sum_{k=0}^\infty \frac{t^k}{k!} A_{uu}^k \vec z_0=\sum_{j} (e^{A_{uu} t})_{1j} x_0^{j+1}.\label{eq:walks1d}
\end{eqnarray}
Note that
\begin{eqnarray}
    (A_{uu})_{ij}= (i+1) \delta_{i+1,j} \label{eq:alice}
\end{eqnarray}
Then,
\begin{eqnarray}
    (A_{uu}^r)_{ab}&=&\sum_{k_1\cdots k_{r-1}} (A_{uu})_{a k_1}(A_{uu})_{k_1k_2}\cdots (A_{uu})_{k_{r-1} b}\nonumber \\
    &=&(a+1)(a+2)\cdots (a+r) \delta_{a+r+1,b} \nonumber \\
    &=&\frac{(a+r)!}{a!}\delta_{a+r,b}
\end{eqnarray}
Then, 
\begin{eqnarray}
    (e^{A_{uu} t})_{ab}=\sum_{r=0}^\infty \frac{t^r}{r!} (A_{uu}^r)_{ab}=\sum_{r=0}^\infty \frac{(a+r)! t^r}{a!r!}\delta_{a+r,b}
\end{eqnarray}
from which we get then
\begin{eqnarray}
    x(t)&=&x_0+\int^t ds \sum_{r=0}^\infty \frac{(r+1)! s^r}{1!r!}\sum_{b=1}^\infty \delta_{r+1,b} x_0^{b+1} \\
    &=&x_0+\sum_{r=0}^\infty \frac{(r+1)! t^{r+1}}{r!(r+1)} x_0^{r+2} \\
    &=&\sum_{r=0}^\infty x_0^{r+1} t^{r}=x_0\sum_{r=0}^\infty x_0^{r} t^{r}=\frac{x_0}{1- x_0 t}.
\end{eqnarray}
which is exactly the solution of $x'=x^2$ with $x(0)=x_0$.
The solution above was derived already in \cite{linkoop} along similar lines. We show now that there is a completely different approach to obtaining the solution.

\begin{figure}[!t]
    \centering
    \includegraphics[width=0.65\textwidth]{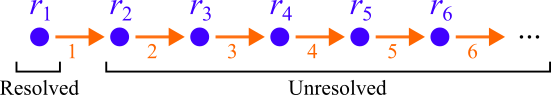}
    \caption{The path graph of the $\dot{x}=x^2$ model}
    \label{fig:pathGraph}
\end{figure}

\subsection{Generating function of directed lattice walks} \label{sec:33}

Here we wish to show that we can express the solution in terms of the generating function of a walk on a path graph. 

We simply define a walk as a weighted sequence of nodes.  To see this, we first introduce the path graph, as illustrated in Fig.~\ref{fig:pathGraph}. The path graph is a directed graph, whose nodes are observables $r_k:=x^{k}$ and weighted edges are the interactions between the observables, i.e., the weight between nodes $r_k$ and $r_{k+1}$ is $k$ (because $\dot{r_k} = k r_{k+1}$). We further define $\mathcal N^k_{(i,j)}$, $i,j \in \mathbb{N}$, as the product of weights along the (only) path of length $k$ connecting nodes $r_i$ and $r_j$. However, since our graph is directed, we must have $\mathcal N^{k}_{(i,j)}=\delta_{k,|i-j|} \mathcal M (i,j)$, where
\begin{equation}
    \mathcal M(i,j):=\frac{(j-1)!}{(i-1)!}. \label{eq:pathWeights}
\end{equation} 

We now make a connection between the path graph and the key term in Eq.~\eqref{eq:walks1d}, $\left(A_{uu}\right)^k_{1,j}$:
\begin{equation}
    \left(A_{uu}\right)^k_{1,j} = \sum_{i_1} \ldots \sum_{i_k} \left(A_{uu}\right)_{1,i_1} \ldots \left(A_{uu}\right)_{i_k,j}.
\end{equation}
Because of the super-diagonal structure of the matrix $A_{uu}$ (see Eq.~\eqref{eq:alice}), each of the $A_{uu}$ in the above product can be interpreted as a weight of a single step moving from a node $r_j$ to the next $r_{j+1}$. As such, the whole sum can be interpreted as the product of the weights along a path with length $k$ connecting nodes $r_2$ and $r_{j+1}$ (note that the indices of unresolved observables begin with $2$:
\begin{equation}
    \left(A_{uu}\right)^k_{1,j} = \delta_{k, j-1} \mathcal M(2,j+1)=\delta_{k,j-1} j!. 
\end{equation}
Then, Eq. (\ref{eq:walks1d}) can be expressed straightforwardly
\begin{align}
A_{ru} \sum_{k=0}^\infty \frac{t^k}{k!} A_{uu}^k \vec z_0 {}&= \sum_{k=0}^\infty \sum_{j=1}^\infty \frac{t^k}{k!} \left(A_{uu}^k\right)_j x_0^{j+1} \nonumber \\ {}&= \sum_{k=0}^\infty \sum_{j=1}^\infty \frac{t^k}{k!} \delta_{k,j-1} \mathcal M(2,j+1) x_0^{j+1} = \sum_{k=0}^\infty \frac{t^k}{k!} \mathcal M(2,k+2) x_0^{k+2}. 
\end{align}

We now define a generating function \cite{wilf2005} for the weighted walks on the path graph:
\begin{equation}
    G(s,x):= \sum_{k=0}^\infty \frac{s^{k} x^{k+2}}{k!} \mathcal M(2, k+2), \label{eq:generatingFunctionx2}
 \end{equation}
which can be calculated analytically, and we explicitly computed it in Sec.~\ref{sec:32}. One can see that the Taylor expansion of $G$ in the parameter $s$ controls the length of the walk, while the Taylor expansion in $x$ controls between which nodes of the graph the walk occurs, as the terms monomials $s^k x^{k+2}$ multiply the number of walks $\mathcal N(2,k+2)$ between node $2$ and $2+k$.   Plugging these in the definition Eq.~\eqref{eq:pathWeights}, we obtain:
\begin{equation}
 G(s,x) = x^2 \sum_{k=0}^\infty \frac{(sx)^k}{k!} (k+1)! =x^2 \sum_{k=0}^\infty (k+1) (sx)^k = \frac{x^2}{(1-sx)^2}. 
\end{equation}

We can then interpret the solution of the differential equation as the time integrand of the generating function of these weighted walks evaluated at $x=x_0$:
\begin{eqnarray}
    x(t)&=&x_0+\int_0^t G(x_0,s) ds \label{eq:genfun1} \\
    &=&x_0+\frac{t x_0^2}{1-t x_0}=\frac{x_0}{1-t x_0}
\end{eqnarray}
which is the expression we had obtained before, but interpreted now in terms of the generating function. 

\subsection{Closed form PDE for the generating function} \label{sec:34}

We now establish that the generating function satisfies a Partial Differential Equation (PDE), which is what we will use for the Lotka--Volterra equation below. Applying $\partial_x$ and $\partial_s$ to the generating function Eq.~\eqref{eq:generatingFunctionx2}, we obtain:
\begin{align}
    \partial_x G(s,x) ={}&  x  \sum_{k=0}^{\infty} (k+2) \frac{t^{k} x^{k}}{(k-1)!} \mathcal M(2,k+2) = x^3 (k+2) \sum_{k=0}^{\infty} \frac{t^{k} x^{k}}{k!} \mathcal M(2,k+2), \\
    \partial_s G(s,x) ={}& x^2 \sum_{k=1}^{\infty} \frac{t^{k-1} x^{k}}{(k-1)!} \mathcal M(2,k+2) = x^3  \sum_{k=0}^{\infty} (k+2) \frac{t^{k} x^{k}}{k!} \mathcal M(2,k+2).
\end{align}
In the derivation above, we have used the identity 
\begin{align}
   \mathcal M(j,k+3) = \left(k+2\right) \mathcal M(j, k+2). 
\end{align}
Clearly, the above equations suggest the closed-form PDE
\begin{eqnarray}
\partial_t\big(G(x,t)\big)=x^2\partial_x  G(x,t).
\end{eqnarray}
The solution of the differential equation can be derived using the method of characteristics. 
We impose the Lagrange-Charpit equality
\begin{eqnarray}
    \frac{dt}{-1}=\frac{dx}{x^2}=\frac{dG}{0}
\end{eqnarray}
From the first two, we get
\begin{eqnarray}
    -t+\Phi=-1/x.
\end{eqnarray}
From the latter, we get $dG=0\rightarrow G=\Gamma$. Imposing the Cauchy surface condition that $\Gamma$ depends on $\Phi$, $G=\Gamma(\Phi)=f(\frac{1-tx}{x})$.
Thus, $G(x,t)=f(\frac{1-t x}{x})$.  We know however that $G(x,0)=x^2$, and thus $f(\cdot)=(\cdot)^{-2}$. Then, $G(x,t)=x^2/(1-x t)^2$ which is the function we found earlier while summing the series explicitly. Then, the interpretation of the solution of Eq.~\eqref{eq:genfun1} can also be applied to the case of other nonlinearly coupled equations.

\subsection{Connection to the Koopman's representation}\label{sec:koopmanx2}

The formal solution Eq.~\eqref{eq:genfun1} can be derived from yet another method, which is arguably the most formal and general derivation. Such a derivation is tightly connected to Koopman's representation \cite{Koopman255,Koopman315}, with which one aims to prescribe the evolutionary equation of the observables. Let us consider a general observable $g:\mathbb{R}^1\rightarrow \mathbb{R}^1$. In the Koopman picture, the states do not move and stay at $x_0$, the initial condition, but the observable $g$ becomes time-dependent, denoted by $g_t:=\mathcal{K}_t g$ where $\mathcal{K}_t$ is the finite-time Koopman operator. We write a field $\psi(x,t)$ satisfying the following equation
\begin{subequations}\label{eq:backwardPDE}
\begin{align}
    \partial_t \psi(x,t)  ={}&  \mathcal{L} \psi(x,t), \\
    \psi(x,0) ={}& g(x),
\end{align}
\end{subequations}
where $\mathcal{L}$ is the (backward) generator of the process, $\mathcal{L}=x^2 \partial_x$. The above PDE and the associated initial condition are often referred to as the Liouville equation \cite{chorin00optimal,AlexandreJ.Chorin2002}, albeit this nomenclature is not standard in non-equilibrium statistical physics\footnote{In statistical physics, the Liouville equation is the forward equation of the dynamics, where the state variables are time-dependent and the observables are static, which is the adjoint of Eq.~\eqref{eq:backwardPDE}}. We note that the structure of the PDE is exactly the one derived earlier.

The solution  of
\begin{equation}
    \psi(x_0,t) = \left(\mathcal{K}_t g \right) (x_0)  \equiv g\left( x(t;x_0) \right)
\end{equation}
contains all the information of the ODE system, with an arbitrary $g$. For example, the formal state solution $x(t)$ can be obtained by setting $g(x):=x^2$, as will be seen below. One can also consider the indicator function $g(x):=\delta(x-x')$ to probe if the solution is at a query point $x'$ at an arbitrary time $t>0$. 

Interestingly, from this Koopman viewpoint, the PDE above is often solved implicitly in terms of the ODE solution. In general, solving the backward PDE \eqref{eq:backwardPDE} is a challenging task. For the $\dot{x}=x^2$ model, we have the ODE solution 
\begin{equation}
    x(t) = \frac{x_0}{1-x_0 t}.
\end{equation}
In line with the method of characteristics, we can ``pull-back'' by express the $x(t)$ as a function of the initial condition $x_0$:
\begin{equation}
    \psi(x,t) = g\left( \frac{x }{1 - t x}\right).
\end{equation}
One can show that the PDE is satisfied:
\begin{subequations}
\begin{align}
    \partial_t \psi(x,t) = {}&  g' \left( \frac{x }{1 -  t x}\right) \frac{x^2}{\left(1 -  t x\right)^2} \\
     x^2 \partial_x \psi(x,t) ={}&   x^2 g' \left( \frac{x }{1 -  t x}\right) \left[ \frac{1}{1-tx} +  \frac{x t }{\left(1+tx\right)^2}  \right] \\
    =&   g' \left( \frac{x }{1 -  t x}\right) \frac{x^2}{\left(1 -  t x\right)^2}.
\end{align}
\end{subequations}

To connect to the solution in the previous calculation, we note that the formal solution of the ODE can be written as
\begin{equation}
    x(t;x_0) = x_0 + \int_0^t x^2\left(s\right)\, ds,
\end{equation}
With this expression, it is clear that the observable of interest should be $g(x):=x^2 $, leading to the solution 
\begin{equation}
    \psi(x,t) = g\left( \frac{x }{1 - t x}\right) =  \left( \frac{x }{1 - t x}\right)^2,
\end{equation}
which is exactly the generating function Eq.~\eqref{eq:generatingFunctionx2}. 

\subsection{Properties of the formal solution} \label{sec:35}
We note that this approach allows us to write explicitly the 
properties of the solution as a function of the initial conditions. For instance, for the example above, we have
\begin{eqnarray}
    \partial_{x_0}x(t)&=&1+\int_0^t \partial_{x_0} G(x_0,t) dt \label{eq:formsol} \\
    &=&1+\frac{1}{x_0^2}\int_0^t \partial_t G(x_0,t) dt \nonumber \\
    &=&1+\frac{G(x_0,t)-G(x_0,0)}{x_0^2} \nonumber \\
    &=&1+\frac{G(x_0,t)-x_0^2}{x_0^2}=\frac{G(x_0,t)}{x_0^2}
\end{eqnarray}
which can be promptly checked using the solution.
It is also interesting to note that there is another way of deriving this PDE, knowing the expression for the solution. Using Eqn.~\eqref{eq:formsol}, and from the fundamental theorem of calculus, we can equate
\begin{eqnarray}
    x^2=x'(t)=G(x,t)
\end{eqnarray}
which we note to be also satisfied. We know from the exact solution that $G(x,t)=x_0^2/(1-x_0 t)^2=x(t)^2$. Now note that from the expression above, we can obtain 
\begin{eqnarray}
    \partial_t G=2 x x'=2 x^3=x^2 \partial_x G,
\end{eqnarray}
which is exactly the PDE we derived using the generating function method.

We wish to show that a similar approach is also possible for the Lotka-Volterra equations, where however the combinatorial structure is slightly more involved, and where the solution is only formal.

\section{The Lotka-Volterra equations} \label{sec:4}
After having discussed at length the techniques used for the simpler toy model of eqn. (\ref{eq:n1d}), let us consider the case of the Lotka-Volterra equation (LV). The solution of the quadratic case shares in fact many similarities with the formal solution we obtain below.

First, we begin by writing the LV equations in the form
\begin{eqnarray}
        \frac{dx}{dt}&=&\alpha x +\beta xy\equiv f(x,y)
        \nonumber \\
        \frac{dy}{dt}&=&\gamma y
        +\delta xy \equiv g(x,y).
\end{eqnarray}
In the typical formulation of the LVs, $\alpha>0$, $\beta<0$, $\gamma<0$, and $\delta>0$.

\subsection{Carleman linearization of LV} \label{sec:41}
As for the case of the toy model, we perform a Carleman linearization, where now the variables are the following monomials
\begin{eqnarray}
    r_{(a,b)}= x^a y^b.
\end{eqnarray}
We have then
\begin{eqnarray}
    \frac{dr_{(a,b)}}{dt}&=&a x^{a-1}(\beta xy+\alpha x)y^b +b y^{b-1}(\delta xy+\gamma y)x^{a}\nonumber  \\
&=&b \delta  x^{a+1} y^b+a \beta  x^a y^{b+1}+x^a y^b (a \alpha +b \gamma )\nonumber \\
&=&a\beta r_{(a,b+1)}+b \delta r_{(a+1,b)} +(a \alpha +b \gamma ) r_{(a,b)}
\end{eqnarray}
then, if we consider the combined index $(a,b)$, 
it follows that the Lotka-Volterra equations,
the linearized LV equation becomes then
\begin{eqnarray}
    \frac{dr_{(0,1)}}{dt}&=& \delta r_{(1,1)}+\gamma r_{(0,1)}\nonumber \\
    \frac{dr_{(1,0)}}{dt}&=&\beta r_{(1,1)}+\alpha r_{(1,0)}\nonumber \\
    \frac{dr_{(1,1)}}{dt}&=&\beta r_{(1,2)}+ \delta r_{(2,1)} +( \alpha + \gamma ) r_{(1,1)}\nonumber \\
    \frac{dr_{(1,2)}}{dt}&=&\beta r_{(1,3)}+ 2\delta r_{(2,2)} +( \alpha + 2\gamma ) r_{(1,2)}\nonumber \\
    \frac{dr_{(2,1)}}{dt}&=&2\beta r_{(2,2)}+ \delta r_{(3,1)} +( 2\alpha + \gamma ) r_{(2,1)}\nonumber \\
    &\vdots&
\end{eqnarray}
It is important to stress that the one above is an \textit{exact} representation of the Lotka-Volterra system. The LV equations can represent a path graph, which is the directed two-dimensional lattice shown in Fig.~\ref{fig:lattice}. 
\subsection{Mori-Zwanzig reduction} \label{sec:42}
\begin{figure}
    \centering
    \includegraphics[scale=0.4]{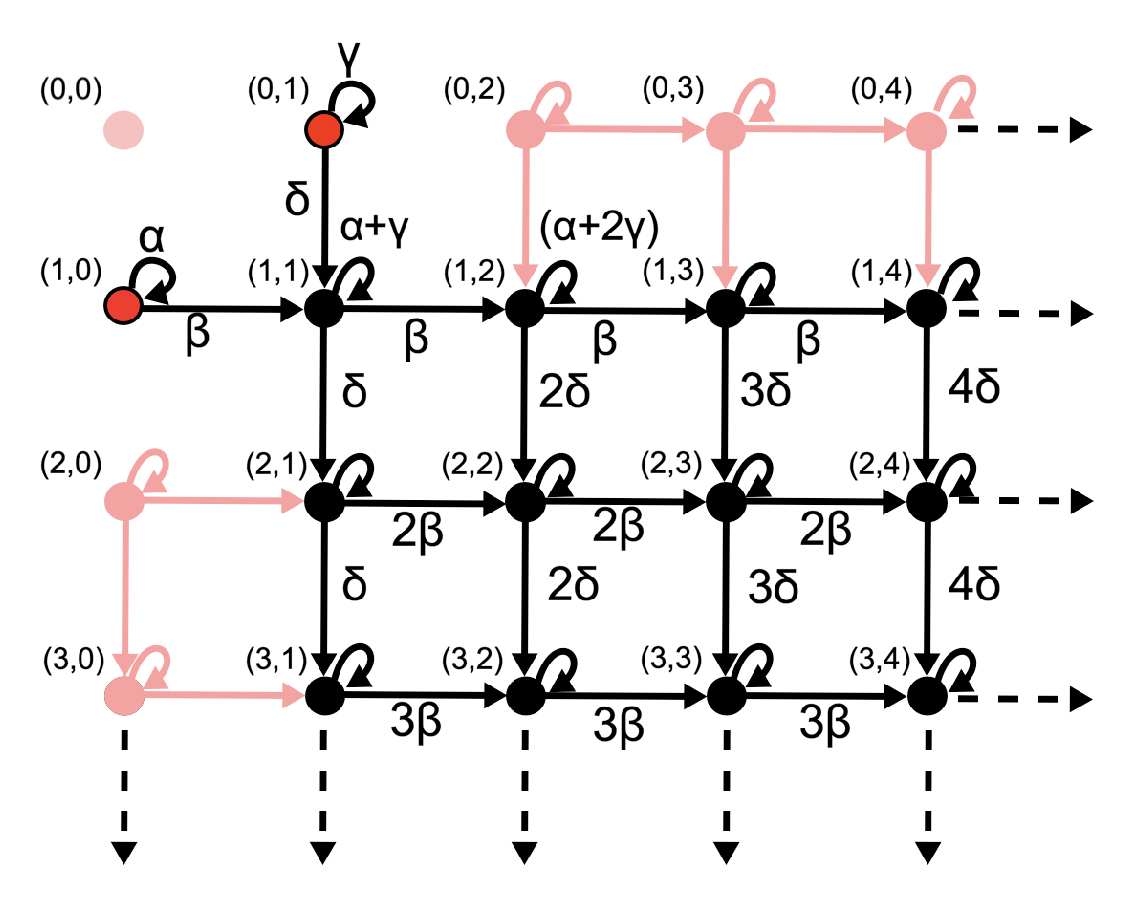}
    \caption{The matrix $A$ from the Carleman linearization in transition between the variables $r_{(a,b)}$ for a portion of the wedge containing the variables $x=r_{(1,0)}$ and $y=r_{(0,1)}$. The variable $r_{(0,0)}$ are constants and thus $dr_{(0,0)}/dt=0$. The pink variables are not involved in the lattice calculations.}
    \label{fig:lattice}
\end{figure}

Because of the linear relationship between the variables, we can write the equations in the form
\begin{eqnarray}
    \frac{d}{dt} \vec r=A \vec r,
\end{eqnarray}
where the vector $(\vec r)_{(a,b)}(t)=x^a(t) y^b(t)$, and
where the matrix $A$ is an upper triangular matrix that can be expressed as a weighted directed graph like the one shown in Fig. 
\ref{fig:lattice}. Let us call $\vec v=(r_{(1,0)},r_{(0,1)})\equiv (x,y)$ and $\vec z=(r_{(1,1)},r_{(2,1)},r_{(1,2)},\cdots)$, so that $\vec r^t=(\vec v^t,\vec z^t)$.
Using equation (\ref{eq:morizwanziglin}), we have
\begin{eqnarray}
    \frac{d\vec z}{dt}&=&A_{ru} e^{A_{uu} t} \vec z 
\end{eqnarray}
which is decoupled from the variables $r_{(1,0)}$ and $r_{(0,1)}$.

The matrix $A$, whose first two rows and columns correspond
to $r_{(0,1)}$ and $r_{(1,0)}$, given the ordering $r_{(0,1)}$, $r_{(1,0)}$, $r_{(1,1)}$, $r_{(2,1)}$, $r_{(1,2)}$, can be written in the form:
\begin{eqnarray}
    A=\left[\begin{array}{cc|cccc}
    \gamma & 0 &  \delta& 0 & 0 & \cdots \\
     0 & \alpha &  \beta & 0 & 0 & \cdots\\
     \hline
     0 & 0 & (\alpha+\gamma) & \delta  & \beta & \cdots \\
     0 & 0 & 0 & (\alpha+2\gamma) & 0  &\cdots \\
     0 & 0 & 0 & 0 & (2\alpha+\gamma)  &\cdots \\
     \vdots & \vdots  & \vdots & \vdots  &\vdots &\ddots
    \end{array}\right]\nonumber 
    \label{eq:partm}
\end{eqnarray}

From the structure of the matrix above, we see that the solution can be written in the form
\begin{eqnarray}
    \frac{d\vec v}{dt}&=& A_{rr} \vec v+A_{ru} e^{A_{uu} t} \vec z_0 \label{eq:morizwanziglinlv}\nonumber 
\end{eqnarray}
The equation above is a linear system of 2 equations with a non-homogeneous forcing.
The set of linearly coupled differential equations above can be solved analytically, with a solution of the form
\begin{equation}
    \vec v(t)=e^{A_{rr}t} \vec v_0+e^{A_{rr}t}\int_0^t ds\ e^{ -s A_{rr}}A_{ru} e^{A_{uu} s} \vec z_0 \label{eq:inte1}.
\end{equation}

The simplicity of the solution is only apparent. While $A_{rr}$ is a two-dimensional diagonal matrix, $A_{uu}$ is a semi-infinite matrix. While semi-infinite, the elements of the matrix $(e^{A_{uu} s})_{(a,b)(c,d)}$ can be calculated combinatorially, analyzing $e^{A_{uu} s}=\sum_{k=0}^\infty (A_{uu})^k \frac{s^k}{k!}$. We first discuss the interpretation of this matrix.

\subsection{Directed walks on lattice} \label{sec:43}
Let us then consider $(A_{uu})^k_{(a,b),(c,d)}$. Since the graph representing $A$ is directed, we must have $c>a\geq 0$, $d>b\geq 0$ in our notation. In particular, we must have $|a-c|+|b-d|\leq k$. If $\alpha=\gamma=0$, then the inequality becomes an inequality, and this becomes a problem of counting paths on the lattice from $(a,b)$ to $(c,d)$. Let us now formalize the combinatorics of such a process. We call $\mathcal R=c-a$ and $\mathcal D=d-b$.
Then, in order for a directed walk to start from $(a,b)$ and reach $(c,d)$ in $k$ steps, no matter the order, if we go down $\mathcal D$ and right  $\mathcal R$ times, we will definitely get to $(c,d)$. We can stop anytime then, for a total amount of $k-\mathcal U-\mathcal D=\mathcal S$ times, for a walk of $k$ steps. Walks can then be mapped as permutations of a vector of moves Down, Right, and Stop of the form
\begin{eqnarray}
    \left( \underbrace{D \cdots D}_{\mathcal D\ \textit{times}} \underbrace{R\cdots R}_{\mathcal R\ \text{times}} \underbrace{S\cdots S}_{\mathcal S\ \text{times}} \right)\label{eq:combd}
\end{eqnarray}
Given a configuration, and depending on the edges it goes through, then we will have a numerical coefficient $C_\sigma$ which depends on the permutation\footnote{In fact, in this representation, one has to divide by a symmetry factor corresponding to the number permutations corresponding to the same path. If for instance, we have $(DDRRSS)$ the sequence with which we perform the first or second down move is actually independent. Thus, we must divide a path like the one above by $1/2^3.$}, a coefficient $\beta ^{\mathcal R} \delta^{\mathcal D}$ and a coefficient $\rho$ which depends on the $\alpha$ and $\gamma$ values and product of the weights along the path. We note that in the limit $\alpha,\gamma\rightarrow 0$, $\rho$ becomes $1$ if $\mathcal R+\mathcal D=k$ otherwise, e.g. the walk has to start from $(a,b)$ and reach the final $(c,d)$ point in exactly $k$ moves.
We can then write
\begin{equation}
    (A^k_{22})_{(a,b)(c,d)}
    =\sum_{\sigma \in \mathcal S_k} C^k_{(c,d)}(\sigma) \beta ^{c-a} \delta^{d-b} \rho(\sigma)\theta_k(\sigma)(c-a)\theta(d-b) \label{eq:a22pow}
\end{equation}
where is the Heaviside function, defined as $\theta(x)=1$ for $x\geq 0$ and zero otherwise and $\mathcal S_k$ is the group of permutations of $k$ elements.
We then have
\begin{equation}
    (e^{A_{uu} s})_{(a,b)(c,d)}
    =\sum_{k=0}^\infty \sum_{\sigma \in \mathcal S_k} C^k_{(c,d)}(\sigma) \beta ^{c-a} \delta^{d-b} \rho_k(\sigma)\theta(c-a)\theta(d-b) \frac{s^k}{k!}.
\end{equation}
Let us now write eqn. (\ref{eq:inte1}) explicitly. This is a 2-component equation, while $\vec z_0=(x_0y_0,x_0^2 y_0,x_0 y_0^2,\cdots)$ is the vector of the appropriate powers of initial conditions.
First, we note that $e^{-s A_{rr}}$ is a diagonal matrix with elements $e^{-s \gamma}$, $e^{-s \alpha}$, corresponding to $(0,1)$, $(1,0)$. Similarly, $A_{ru} e^{A_{uu} s}$ selects only the first two rows of $e^{A_{uu} s}$ with the appropriate weights. 
Then, we have
\begin{equation}
    e^{-s A_{rr}} A_{ru} e^{A_{uu} s}\vec z_0=\sum_{c,d}\begin{pmatrix}
         \delta e^{-s \gamma} (e^{A_{uu} s})_{(1,1),(c,d)} (\vec z_{0})_{(c,d)} \\
        \beta e^{-s \alpha} (e^{A_{uu} s})_{(1,1),(c,d)} (\vec z_{0})_{(c,d)} \\
    \end{pmatrix}\label{eq:inte2}
\end{equation}
Then, if we define the diagonal matrix $\text{diag}(\delta e^{-s \gamma},\beta e^{-s \alpha})$ and
\begin{equation}
\eta(s,x_0,y_0)=\sum_{c,d} (e^{A_{uu} s})_{(1,1),(c,d)} (\vec z_{0})_{(c,d)}    \label{eq:eta}
\end{equation}
we get
\begin{equation}
    e^{-s A_{rr}} A_{ur} e^{A_{uu} s}\vec z_0=\eta(s) \begin{pmatrix}\delta e^{-s \gamma} & 0 \\ 0 & \beta e^{-s \alpha} \end{pmatrix} \vec 1.
\end{equation}
and thus the solution can be written as
\begin{eqnarray}
    \begin{pmatrix}
        y(t) \\
        x(t)
    \end{pmatrix}=    \begin{pmatrix}
        e^{\gamma t}y_0 \\
        e^{\alpha t} x_0
    \end{pmatrix}+\int_0^t ds\ \eta(s,x_0,y_0)\begin{pmatrix}
          \delta e^{\gamma (t-s)} \\
          \beta e^{\alpha(t-s)}
    \end{pmatrix}\label{eq:solution}
\end{eqnarray}
We can see from the equation above that for $t=0$ we recover the initial conditions. 

From a perturbative standpoint, we see that in the equation above 
we need to calculate integrals of the form
\begin{eqnarray}
    I_m(t)=\int_0^t \eta(s,x_0,y_0) e^{-m s} ds
\end{eqnarray}
for arbitrary $m$.
If we write
\begin{eqnarray}
\eta(s)=\sum_{c,d\geq 1} \sum_{k=0}^\infty \frac{s^k}{k!}\sum_{\sigma \in \mathcal S_k} C^k_{(c,d)}(\sigma) \beta ^{c-1} \delta^{d-1} \rho_{k}(\sigma) (\vec z_{0})_{(c,d)}    .\label{eq:eta2}
\end{eqnarray}
We see that the expansion can be reduced to expressions of the form
\begin{eqnarray}
    I^k_m(t)=\int_0^t s^k e^{-m s} ds. \label{eq:directIntegral}
\end{eqnarray}
Furthermore, $I^k_m(t)$ can be expressed as
\begin{equation}
    I^k_m(t) = \frac{\Gamma (k+1)-\Gamma (k+1,m t)}{m^{k+1}}, \text{ if } m>0, \label{eq:gammafun}
\end{equation}
where $\Gamma(k+1)=(k+1)!$ and $\Gamma(k+1,mt)=\int_{t}^\infty e^{-s} s^k\ ds$ is the incomplete Gamma function.
Note that for small values of $t$, $I^k_m(t\rightarrow 0^+)={(mt)^k}/{(1+k)}+O(t^{k+1})$, and thus at small values of $t$ the expansion also corresponds to a time Taylor expansion.
Instead, in the limit $m\rightarrow 0$, we have from the properties of the $\Gamma$ functions that
\textit{\begin{eqnarray}
    \lim_{m\rightarrow 0} I_{m}^k(t)=\frac{t^{1+k}}{1+k}.
\end{eqnarray}}
which is consistent with setting $m=0$ in eqn. (\ref{eq:gammafun}) and integrating $s^k$.

The summation over $c$ and $d$ can be swapped with the summation over $k$ and $\sigma$, 
from which we get
\begin{eqnarray}
    \Gamma_k=\sum_{c=1}^\infty  \sum_{d=1}^\infty \sum_{\sigma\in \mathcal S_k}\delta^{d-1} y_0^d \beta^{c-1} x_0^c C^k_{(c,d)}(\sigma) \rho_k(\sigma)
\end{eqnarray}
from which we can write
\begin{eqnarray}
        y(t) &=&e^{\gamma t}y_0+\delta e^{\gamma t} \Omega_\gamma(t) \label{eq:expressionf1}\\
        x(t)&=& e^{\alpha t} x_0+\beta e^{\alpha t} \Omega_\alpha(t)\label{eq:expressionf2}
\end{eqnarray}
where 
\begin{equation}
    \Omega_m(t)=
 \sum_{k=0}^\infty \Gamma_k  I^k_m(t)  \label{eq:omega}
\end{equation}
which is the final expression for the lattice paths expansion solution of the Lotka-Volterra equation. We define the parameters $\Gamma_k$ as the lattice coefficients. However, we wish to stress that Eq.~\eqref{eq:gammafun} is true only for $m>0$. Because conventional parametrization demands $\gamma<0$, one must rely on Eq.~\eqref{eq:directIntegral} to evaluation of $I^k_{\gamma}(t)$. It is clear that $I^k_{\gamma}(t)$ is not converging for larger values of $t$ for any $k\in \mathbb{Z}_{\ge 0}$. Consequently, this formal method by lattice coefficients is unfortunately of little use, if not only formally. 

This is now a good moment to stop and focus on the meaning of these expressions. Equation (\ref{eq:solution}), albeit formal, provides an interpretation of the exact solution of the 2-species Lotka-Volterra equations in terms of directed, and weighted, walks on lattice paths. As a result, the function $\eta(s,x,y)$ can in principle be estimated via a Monte Carlo approach, with random lattice walks stopping at $(a,b)$ in order to estimate the contributions of order $x_0^a y_0^a$ in the initial condition. This is the first result of this paper. Yet, admittedly this is a quite cumbersome approach. Although in the next section, we provide examples for the calculation of the lattice coefficients, it is worth mentioning that the purpose of the rest of the paper will be to show that $\eta$ is the solution of a particular differential equation.

\subsection{Example of lattice coefficient calculation} \label{sec:44}
The solutions of eqn. (\ref{eq:expressionf1})-(\ref{eq:expressionf2}) are, as a matter of fact, written explicitly in terms of coefficients that need to be evaluated at every order $k$ of the lattice expansion. However, unless an exact expression for $\Omega_{m}(t)$ can be found, this is only formal for the time being. In fact, the symmetric group  $\mathcal S_k$ on a set of $k$  elements has order  $k!$. It is then immediate to see that the complexity of each order increases dramatically. However, we still think that this approach can lead to a way of formalizing this solution. 

To show how this expansion works, let us consider the first few orders. We define the boundary set $\mathcal B_k$ of the nodes in the lattice that can be reached in $k$ steps.
 For $k=0$, we identify $\mathcal S_0=\emptyset$.
 In this case $\Gamma_0=r_{(1,1)}=x_0 y_0$. At $k=1$, $\mathcal S_1=\{Id\}$, the boundary of nodes that can be reached after one step is composed of $\mathcal B_1=\{r_{(1,2)},r_{(2,1)}\}$. It is then easy to see that $\sigma=Id$, $\Gamma_1=\beta x_0 y_0^2+\delta x_0^2 y_0$. 
 The first non-trivial case is $k=2$. The boundary of the nodes that can be reached is $\mathcal B_2=\{r_{(1,2)},r_{(2,1)},r_{(2,2)}\}$. The set $\mathcal S_2=\{\sigma_1=\{1,2\},\sigma_2=\{2,1\}\}$.
 
 For the nodes $r_{(1,2)},r_{(2,1)}$ the walk has to stop once at least and go right and down respectively, while for $r_{(2,2)}$ once right and down at least. 
 
\begin{figure}[h!]
   \centering
    \includegraphics[scale=0.25]{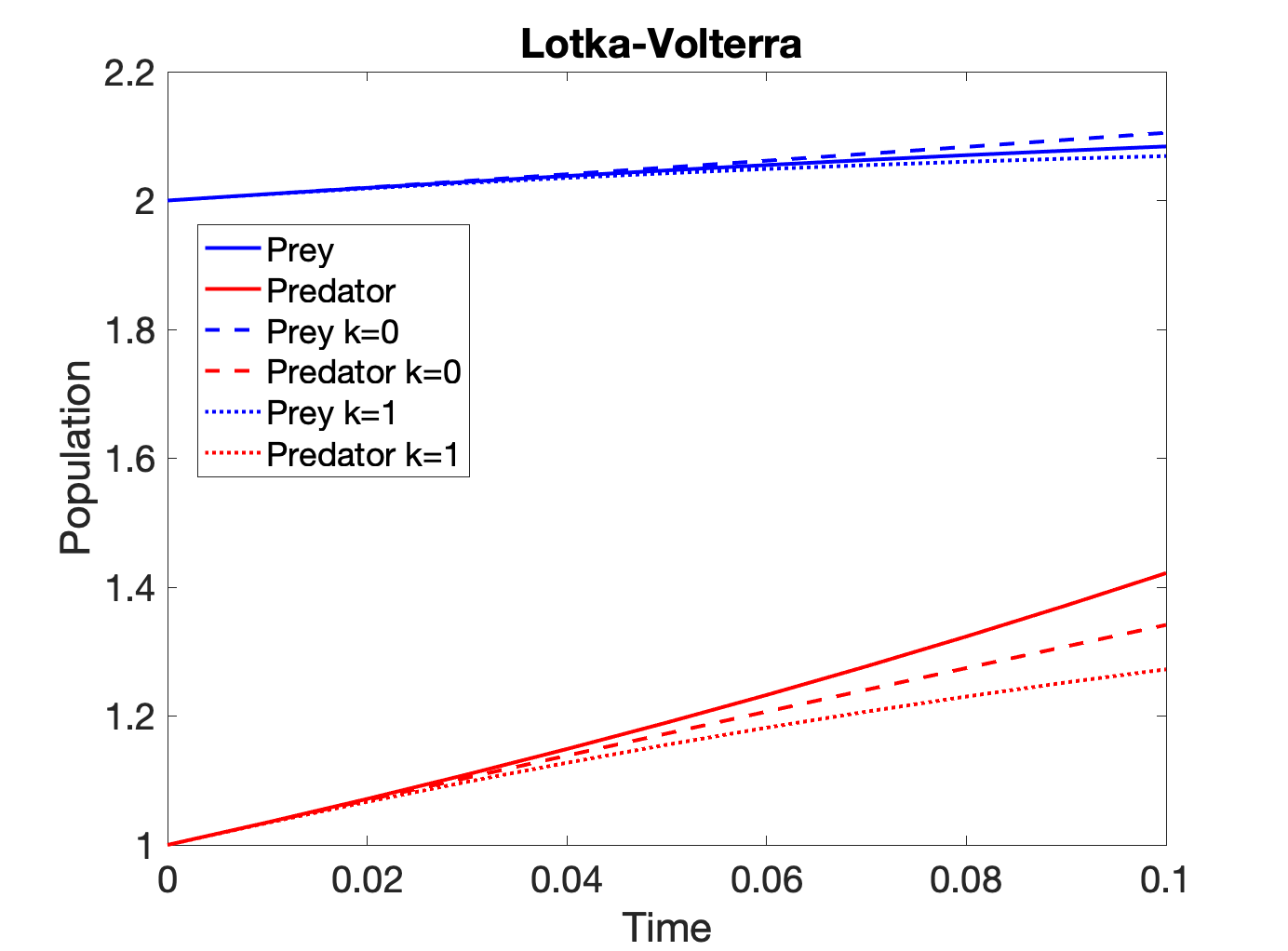}
    \caption{Analytical at order $k$ versus forward Euler method with $dt=0.01$ (solid lines), for the parameters $\alpha = 1$,  
$\beta = -0.5$, 
$\gamma = -0.5$, 
$\delta = 2$. We see that the expansion captures the short time behavior of the system.}
    \label{fig:example}
\end{figure}

 Thus, the transitions are given by
 \begin{enumerate}
 \item for $r_{(1,2)}=x_0 y_0^2$, our moves must be $\sigma_1(R,S)=(R,S)$ or $\sigma_2(R,S)=(S,R)$, thus $C_2(\sigma_1)=C_2(\sigma_2)=1$,  while $\rho_2(\sigma_1)=(\alpha+\gamma)$, $\rho_2(\sigma_2)=(\alpha+2\gamma)$, and $\beta^{c-1}\delta^{d-1}=\beta$;
 \item for $r_{(2,1)}=x_0^2 y_0$ we have $\sigma_1(D,S)=\{D,S\}$ and $\sigma_2(D,S)=(S,D)$, thus $C_2(\sigma_1)=1=C_2(\sigma_2)=1$, while $\rho_2(\sigma_1)=\alpha+\gamma$ and $\rho_2(\sigma_2)=2\alpha+\gamma$, with  $\beta^{c-1}\delta^{d-1}=\delta$;
 \item for $r_{(2,2)}$ we must have $\sigma_1(R,D)=(R,D)$ and $\sigma_2(R,D)=(D,R)$, and $C_2(\sigma_1)=C_2(\sigma_2)=2$, and $\rho_2=1$ and $\beta^{c-1}\delta^{d-1}=\beta\delta$.
\end{enumerate}
These correspond then to 
\begin{eqnarray}
\Gamma_2&=&\beta\big((\alpha+\gamma)+(\alpha+2\gamma)\big)x_0 y_0^2\nonumber \\
&&+\delta\big((\alpha+\gamma)+(2\alpha+\gamma)\big)x_0^2 y_0 \nonumber \\
&&+\big((\beta)(2\delta)+(2\beta)(\delta) \big)x_0^2 y_0^2 \nonumber \\
&=&\beta (2\alpha+3\gamma) x_0 y_0^2+\delta(3\alpha+2\gamma) x_0^2 y_0+4 \beta \delta x_0^2 y_0^2.\nonumber \\
\end{eqnarray}
We then get that the expansion up to order $k=2$ is given by
\begin{eqnarray}
        y(t) &=&e^{\gamma t}y_0+\delta e^{\gamma t} \Big(x_0y_0\frac{1-e^{-\gamma t}}{\gamma}+\Gamma_1 I_\gamma^1(t)   +\Gamma_2 I_{\gamma}^2(t)\Big) \nonumber \\
        &&+O(t^3)\nonumber \\
        x(t)&=& e^{\alpha t} x_0+\beta e^{\alpha t} \Big(x_0y_0\frac{1-e^{-\alpha t}}{\alpha}+\Gamma_1 I_\alpha^1(t)   +\Gamma_2 I_{\alpha}^2(t)\Big) \nonumber \\
        &&+O(t^3)\nonumber 
\end{eqnarray}
which provides an example of the form of the solution. Then up to order two, the solution above versus the numerically obtained one is shown in Fig. \ref{fig:example}. We see that this expansion is problematic at a perturbative level, because of its slow convergence, and for $\beta<0, \gamma<0$, it has an alternating sign which does not improve the convergence either.
Albeit in the following we focus on deriving an exact equation for $\eta$, the approach described above allows in principle to study the solution of LV equations using Monte Carlo techniques.

\subsection{Lotka-Volterra via generating functions}  \label{sec:45}
The analysis of the previous section has highlighted the importance of deriving exact equations for $\eta$.
As we saw in the plots above, the perturbative expansion is slowly converging. We thus need a better expression for the solution.

We now know however that $\eta(t,x,y)$ in eqn. (\ref{eq:eta}) is exactly the generator function of walks of length $k$ from $(1,1)$ to $(c,d)$, and the coefficient corresponding to $t^k/k^2 (x^c y^d)$ is the weighted walk of length $k$.  Similar to what we introduced earlier for the simpler model in eqn. (\ref{eq:pathWeights}) for the model of eqn. (\ref{eq:n1d}), we now have a weighted walk coefficient $\mathcal N^k_{(x_1,y_1),(x_2,y_2)}$ which is the sum of all possible walks from vertex $(x_1,y_1)$ to vertex $(x_2,y_2)$ on a lattice. In our case, since we are interested in walks of length $k$ starting always from node $(1,1)$, we can simply define $\mathcal N^{k}_{(1,1),(a,b)}=N^k_{(a,b)}$. 

We can then write, similarly to what we did for the simpler ODE of eqn. (\ref{eq:n1d}), a recursion relation of the form
\begin{equation}
    N^k_{(a,b)}=(\alpha a+ b\gamma)N^{k-1}_{(a,b)}+a\beta N^{k-1}_{(a,b-1)}+b\delta N^{k-1}_{(a-1,b)} \label{eq:LVrecursion},
\end{equation}
where $N^k_{a,b}$ is the product of the weights along a path of length $k$ connecting $(1,1)$ and $(a,b)$ (the dependence of $(1,1)$ is suppressed for brevity). 
It is clear that we have the following conditions: (1) $N^0_{(1,1)}=1$, (2) $N^k_{(0,1)}=N^k_{(1,0)}=0$, $k\in \mathbb{Z}_{\ge 0}$, and (3) $N^{r}_{(a,b)} = 0$ if $r \ne a+b-2$, $a,b \in \mathbb{N}$.
Then, the generating function can be written in the form
\begin{eqnarray}
\eta(t,x,y)=\sum_{a,b=1}^\infty\sum_{k=0}^\infty N^k_{(a,b)}  \frac{t^{k}}{k!}x^a y^b.
\end{eqnarray}

We multiply the left and right of the recursion relation Eq.~\eqref{eq:LVrecursion} by $t^{k-1}x^a y^b/{(k-1)!}$ and sum over $a$, $b$, and $k$ from $1$ to $\infty$.
Let us analyze this term by term. On the left-hand side, we obtain $ \sum_{a,b=1}^\infty \sum_{k=1}^\infty t^{k-1} x^a y^b N_{(a,b)}^k / {(k-1)!}=\partial_t \eta(t,x,y).$  On the right hand side, the summation over $k$ is exactly right to give the $t$ dependence, and we focus on the $a,b$ summations. 
\begin{eqnarray}
\alpha:&& \sum_{a,b=1}^\infty\sum_{k=1}^\infty \frac{t^{k-1}}{(k-1)!} a x^a y^b N_{(a,b)}^{k-1} =x \partial_x \eta(t,x,y) \nonumber \\
\gamma:&&\sum_{a,b=1}^\infty \sum_{k=1}^\infty\frac{t^{k-1}}{(k-1)!} b x^a y^b N_{(a,b)}^{k-1} =y \partial_y \eta(t,x,y) \nonumber \\
\delta:&& \sum_{a,b=1}^\infty \sum_{k=1}^\infty\frac{t^{k-1}}{(k-1)!} {b} x^a y^b N_{(a-1,b)}^{k-1} = 
xy\partial_y \eta(t,x,y) \nonumber \\
\beta:&& \sum_{a,b=1}^\infty \sum_{k=1}^\infty\frac{t^{k-1}}{(k-1)!} {a} x^{a} y^{b} N_{(a,b-1)}^{k-1} = x{y}\partial_x   \eta(t,x,y),\nonumber 
\end{eqnarray}
from which we obtain the PDE
\begin{eqnarray}
    \partial_t \eta(t,x,y)&=&\Big((\alpha x +\delta xy)\partial_x +(\gamma y +\beta y x)\partial_y  \Big)\eta(s,x,y) \nonumber  \\
    &=&  \Big(f(x,y)\partial_x +g(x,y)\partial_y\Big)\eta(t,x,y) \\
    \eta(0,x,y)&=&xy\\
    \eta(t,0,y)&=&\eta(t,x,0)=0
\end{eqnarray}
where the boundary conditions are derived as follows. Since $N^k_{(1,0)}=N^r_{(0,1)}=0$, it follows that $\eta(s,0,y)=\eta(s,x,0)=0$. Since $t=0$ must imply that the only surviving term is $N^0_{(1,1)} xy$ with $N^0_{(1,1)}=1$, $\eta(0,x,y)=xy$ must be supplied as initial condition.

The PDE can also be derived from the formal Koopman's picture, analogous to the technique presented in Sec.~\ref{sec:koopmanx2}. To get to the same form as the PDE above, we first rewrite the ODE to absorb the linear terms
\begin{subequations}\label{eq:absorbedLV}
\begin{align}
    \frac{d}{d t} \left( e^{\alpha t} x(t) \right)   ={}&  \beta e^{ \alpha t}  x(t) \, y(t), \\
    \frac{d}{d t} \left( e^{\gamma t} x(t) \right)   ={}& \delta e^{\gamma t}  x(t) \, y(t).
\end{align}
\end{subequations}
To this end, we can see that both terms involve $x(t) y(t)$, hinting that the key observable is $g(x,y):= xy$. As such, the backward PDE (Eq.~\eqref{eq:backwardPDE}) is
\begin{subequations}
\begin{align}
    \partial_t \eta(x,y,t)  ={}&  \left[ \left( \alpha x + \beta x y \right) \partial_x + \left(\gamma y + \delta x y \right) \partial_y \right] \eta(x,t), \\
    \eta(x,y,0) ={}& g(x,y) = xy,
\end{align}    
\end{subequations}
and once $\eta$ is solved (recall that $\eta(x_0,y_0,t) = x(t;x_0)\, y(t;y_0)$, we can integrate Eq.~\eqref{eq:absorbedLV}:
\begin{subequations}\label{eq:LVs2}
\begin{align}
    x(t)  = {}& x(0) + \beta \int_0^t e^{\alpha (t-s) } \eta(x_0, y_0, t) \, d s, \\
    y(t)  = {}& y(0) + \gamma \int_0^t e^{\gamma (t-s) }  \eta(x_0, y_0, t) \, ds.
\end{align}
\end{subequations}
which is exactly the same as the solution obtained from combining Carleman linearization, Mori-Zwanzig formalism, and generating function of the directed weighted graph.

To see that the equation above is correct, let us consider inserting explicitly the formal solution (Eq.~\eqref{eq:solution}) into the ODE. We have
\begin{eqnarray}
    \alpha x+\beta xy&=&x'(t)\nonumber \\
    &=&\alpha e^{\alpha t} x_0+\beta\left[\alpha \int_0^t \eta(x_0,y_0,t) e^{\alpha(t-s)} ds +\eta(x_0,y_0,t) \right]
\end{eqnarray}
We now note that on the right-hand side, we have
\begin{equation}
    \alpha x+\beta xy=x'(t)=\alpha x(t)+\beta \eta(x_0,y_0,t)
\end{equation}
from which we obtain the identity
\begin{equation}
    \eta(x_0,y_0,t)\equiv \eta(x(t),y(t))=x(t)y(t)
\end{equation}
we now take the derivative with respect to time, obtaining
\begin{eqnarray}
    \partial_t \eta(x,y)&=&x'(t)y(t)+x(t)y'(t)\nonumber \\
    &=&(\alpha x+\beta x y)y(t)+(\gamma y+\delta x y)x(t)\nonumber 
\end{eqnarray}
Now note that we can write
\begin{eqnarray}
    x(t)&=&\partial_y \eta\\
    y(t)&=&\partial_x \eta
\end{eqnarray}
From which we obtain
\begin{eqnarray}
\partial_t \eta=(\alpha x+\beta x y)\partial_x \eta+(\gamma y+\delta x y)\partial_y \eta\label{eq:genfun2dlv}
\end{eqnarray}
Which is the expression obtained using the generating function method.

The solution of the PDE above can be solved (formally, and in principle) on a Cauchy surface determined by constants of integration emerging from the Lagrange-Charpit equations, as the above is a quasi-linear equation, exactly as in the case of the simpler example described earlier. However, our attempts to solve the equation analytically lead only to a complicated implicit form that we decided to omit in the present manuscript.

\subsection{Comments on Lotka-Volterra with N species and hyperlattices} \label{sec:46}
Such methodology for the generation of exact solutions of Lotka-Volterra equations can be generalized to higher dimensions. We consider then the set of coupled differential equations of the form
\begin{equation}
    \frac{dx_i}{dt}=\alpha_i x_i+x_i \sum_{j}^N \beta_{ij} x_j
\end{equation}
where the matrix $\beta$ is zero on the diagonal.
Similarly to what we had done in the previous section, we consider the observables $r_{\vec a}=\prod_{i=1}^N x_i^{a_i}$, and focus on their time derivatives
\begin{eqnarray}
    \frac{d}{dt} r_{\vec a}&=&\sum_{j=1}^N a_j x_j^{a_{j-1}}\frac{dx_j}{dt}\prod_{r\neq j} x_r^{a_r}\\
    &=&\sum_{j=1}^N a_j x_j^{a_{j-1}}(\alpha_j x_j+x_j \sum_{k}^N \beta_{jk} x_k)\prod_{r\neq j} x_r^{a_r}\nonumber \\
    &=&  (\sum_{j} a_j \alpha_j) r_{\vec a}+\sum_{j=1}^N a_j \sum_{k}^N \beta_{jk} r_{\vec a+\hat x_k}\nonumber \\
    &=&\tilde \alpha_a  r_{\vec a}+\sum_{k=1}^N \tilde \beta_k r_{\vec a+\hat x_k}.
\end{eqnarray}
As a result, we obtain a lattice equation in higher dimensions. Equation (\ref{eq:combd}) has to be generalized, as now we have $N$ dimensions for the lattice path.  It is known that in general this construction can be expressed in a tensorial representation \cite{pouly}.

The key difference between the $2-$species case and the higher dimensional one is that the lattice depends on the initial point, which makes the analysis slightly more complicated.  For $3-$species, the initial points of the lattice paths are given by $x=r_{(1,0,0)}, y=r_{(0,1,0)}$ and $z=r_{(0,0,1)}$, analogously to $x=r_{(1,0)}$ and $y=r_{(0,1)}$ in Fig. \ref{fig:lattice} in the $2-$species. It can be promptly seen that, since $\dot x$ has the monomials $x$,$xy$ and $xz$ on the right hand side,  $\dot y$ has the monomials $y$,$yx$ and $yz$
and $\dot z$ instead $z$, $zx$ and $zy$, the lattice expansion implies
\begin{eqnarray}
r_{1,0,0}&\rightarrow& r_{1,0,0},r_{1,1,0},r_{1,0,1}\\
r_{0,1,0}&\rightarrow& r_{0,1,0},r_{1,1,0},r_{0,1,1}\\
r_{0,0,1}&\rightarrow& r_{0,0,1},r_{1,0,1},r_{0,1,1}   \label{eq:multi}
\end{eqnarray}
From this we see that the lattice walks for the resolved variables begin at three sublattices starting at $r_{0,1,1}$, $r_{1,0,1}$ and $r_{1,1,0}$ which do not coincide as in the two-dimensional case. For comparison, in the $2-$ species case
we had
\begin{eqnarray}
r_{1,0}&\rightarrow& r_{1,0},r_{1,1}\\
r_{0,1}&\rightarrow& r_{0,1},r_{1,1}
\end{eqnarray}
which implies that we could write the solution in terms of a single generating function $\eta$ corresponding to walks from $r_{1,1}$ to $r_{a,b}$. However, since the lattice paths for a higher number of species are directed, the sublattices resulting from these paths do not coincide in this case. Thus, in the multi-species case, one has to resort to multiple generating functions. In the $3-$species case, these are $\eta_{xy}$, $\eta_{xz}$ and $\eta_{yz}$, related to walks from $r_{1,1,0}$, $r_{1,0,1}$ and $r_{0,1,1}$ to a generic point $r_{a,b,c}$ respectively. Although by symmetry these three must be related, we see immediately that the hyperlattices require a separate discussion that goes beyond the purpose and space limitations of this note. We can see however that our methodology can also be applied here, with due attention.

\section{Conclusions} \label{sec:6}
This paper presents a novel approach to analyzing the Lotka-Volterra equations by introducing a lattice path expansion, preceded by a Carleman linearization and by a Mori-Zwanzig reduction which we used to derive the formal solution. We first analyzed the simpler model $\dot x=x^2$, which although had been solved before, as we have shown her the solution has an interpretation in terms of lattice path expansion. As far as we know, such lattice structure employed in this study was not previously known or utilized in the simpler model examined in the paper, nor had it been applied to the Lotka-Volterra equations.
Our paper elucidates the interplay between the formal solutions of the ODEs, the generating functions of the lattice walks, and how these can be interpreted in terms of Koopman evolution.

This approach provides a powerful method for understanding the Lotka-Volterra equations. Indeed, it came as a posteriori surprise that the generators of these lattice path expansions have also, in turn, an interpretation in terms of Koopman evolution operators, applied to a particular observable. 

Some critical comments are in order. Although we could not provide an exact solution to the two-dimensional Lotka-Volterra equations, what we think to be an important contribution of this paper is the connection between combinatorial methods and the solution of nonlinear ODE, and in turn the relationship between the generator of walks on graphs to the Koopman evolution of observables. We think of the Carleman and Mori-Zwanzig techniques as a method to obtain the formal solution that in principle can be applied, clearly with harder work, to a large class of ODE that can be expressed in terms of monomials of the dynamical variables. The formal solution of the LV equations should be seen only as that, as it is written in terms of the generator of lattice walks on the graph. Solving the PDE of such a generator is as hard, if not more difficult, than solving the ODE directly. In future works we will focus on the numerical analysis of the PDE for the LV equation, trying to find connections between the transition to oscillatory regime in the Lotka-Volterra ODEs and to properties of the related PDE. 

In future work, we will focus on the numerical integration of eqn. (\ref{eq:genfun2dlv}) trying to elucidate these connections.

\backmatter

\bmhead{Acknowledgments}

The work of FC and YTL was carried out under the auspices
of the NNSA of the U.S. DoE at LANL under Contract No. DE-AC52-06NA25396, and in particular support from LDRD via 20220063DR, 20230338ER, and 20230627ER.

\noindent









\bibliography{sn-bibliography}
\end{document}